\documentclass{article}
\usepackage[T1]{fontenc}
\usepackage[utf8]{inputenc}
\usepackage{ismir}
\usepackage{amsmath,cite,url}
\usepackage{graphicx}
\usepackage{color}

\usepackage{etoolbox}

\newcommand{\method}{\textsc{InvFlowFD}\xspace}
\newcommand{\stabilityMethod}{\textsc{StabilityFlow}\xspace}

\usepackage{microtype}
\usepackage{graphicx}
\usepackage{subfigure}
\usepackage{booktabs} 

\usepackage{amsmath}
\usepackage{amssymb}
\usepackage{mathtools}
\usepackage{amsthm}
\usepackage{xspace}

\usepackage{algorithm}

\usepackage{algpseudocode}

\usepackage{titlesec}

\title{\method: Reference-Free and Background-Set-Free Perceptual Music Quality Metric with Flow Matching Inversion}

\oneauthor
  {Alon Ziv \hspace{0.3cm} Harel Pogoda \hspace{0.3cm} Yossi Adi}
  {School of Computer Science and Engineering \\
  The Hebrew University of Jerusalem, Israel\\\\
  \texttt{alonzi@cs.huji.ac.il}}

\def\authorname{A. Ziv, H. Pogoda, and Y. Adi}

\usepackage[bookmarks=false,pdfauthor={\authorname},pdfsubject={\pdfsubject},hidelinks]{hyperref}
\usepackage[capitalize,noabbrev]{cleveref}

\begin{document}

\maketitle

\begin{abstract}
Existing reference-free methods for evaluating music perceptual quality alleviate the need for paired noisy-clean data, but they still rely on a background set, which is used to compute aggregated statistics of clean audio samples. In this work, we propose a novel approach that eliminates this requirement, achieving background-set-free and reference-free quality estimation using only a pre-trained Flow Matching backbone. We demonstrate that unconditional Flow Matching inversion via simple Euler integration is sufficient to detect various artificial distortions and accurately rank music generation models against human perceptual judgments. We introduce \method, which performs flow inversion and compares a group of inverted samples to the prior distribution. We evaluate our method against prior work, quantitatively and with a thorough human study. Results suggest that \method is highly correlated with human perception of sound distortions, as well as generative models' quality, while being more flexible and less restrictive than existing metrics. 
\end{abstract}

\section{Introduction}
\label{sec:introduction}
\begin{figure*}[ht!]
    \hspace*{-1.4cm}
    \centering
    \includegraphics[scale=0.7]{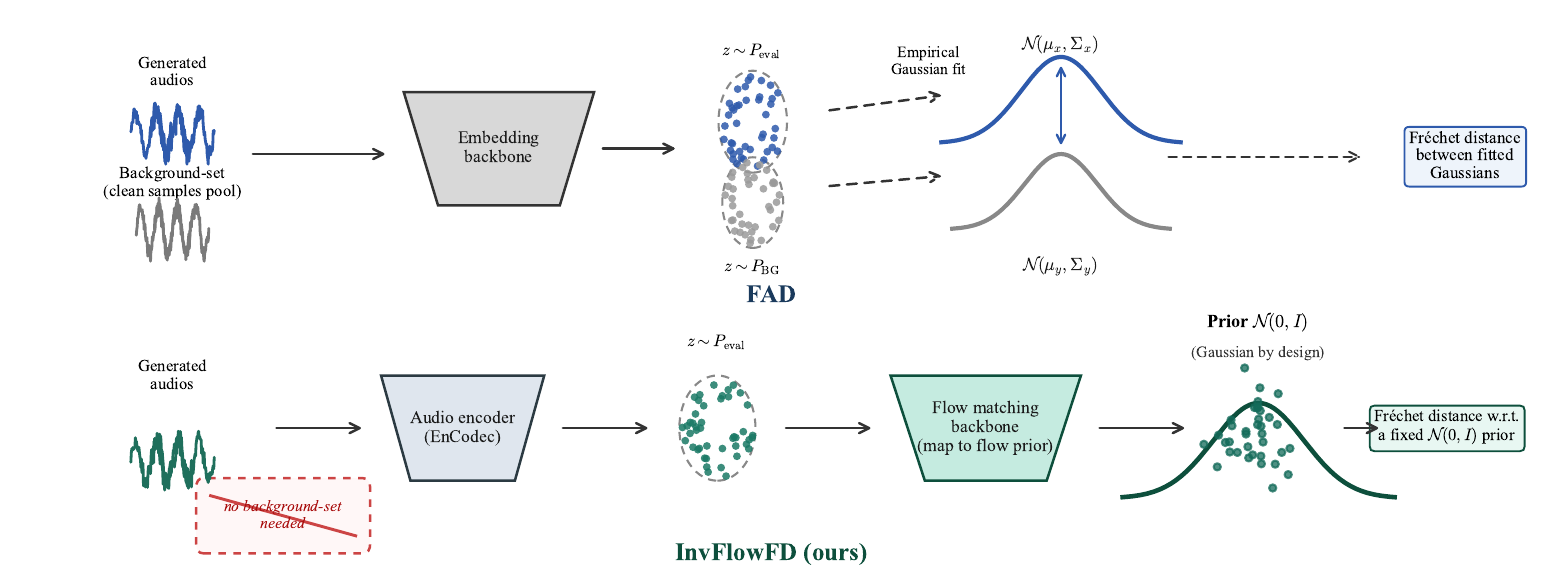} 
    \caption{Overview of our methodology. \method eliminates the need for a background set compared to FAD.\label{fig:inv_flow_all}}
\end{figure*}

Automatically measuring the perceptual quality of music is challenging, and could be done in several different ways. One approach would be to directly fit a predictor to human ratings of quality, as recently done in large scale for the Audiobox Aesthetic predictors \cite{tjandrametaaudioboxaestheticsunified:25}. A second approach, would include an aligned \textbf{\textit{reference set}}, holding the exact corresponding clean signal for each evaluated sample. 

Collecting human ratings might be costly, and having an aligned reference set might be impractical in many real-world scenarios. As an alternative for both, prior work on automatic evaluation of the perceptual quality of music, such as FAD \cite{kilgourfrechetaudiodistancemetric:19}, introduced reference-free methods, alleviating the need for noisy-clean paired data. These methods leveraged a pre-trained neural network \textbf{\textit{backbone}} such as VGGish \cite{hersheycnnarchitectureslargescaleaudio:17} or CLAP \cite{wulargescalecontrastivelanguageaudiopretraining:24} to obtain a reference-free evaluation, through aggregated statistics over clean data. However, these methods still rely on the existence of a large set of studio-recorded samples from which the background statistics are being exacted, namely the \textbf{\textit{background set}}. 
The need to provide a background set, in order to use FAD \cite{kilgourfrechetaudiodistancemetric:19} and other comparison evaluation methods, injects a bias into the evaluation process, as different background sets imply different results. Generally, we would like the evaluated music to be compared to the widest and most evenly distributed dataset could be found. Since it is costly and technically complicated to assemble and use such a ground-truth set, the evaluation process is affected by the chosen background samples. We find that background-set-dependent metrics such as FAD, are often ambiguous, reporting different trends for different background sets. We demonstrate this sensitivity both in terms of the reaction to synthetic distortions, and in terms of the correlation with human perception.

Crucially, this paper demonstrates that music perceptual quality can be measured automatically without having a background set of clean samples at all. Relying only on a learned Flow Matching \cite{lipmanflowmatchinggenerativemodeling:23} backbone, we show how a simple unconditional inversion with Euler integration, is sufficient for detecting various types of artificial distortions, as well as for ranking the quality of generative models.
\textit{\textbf{Our methodology eliminates completely the dependency on a background set and its statistics.}}

We introduce \method, a \textit{background-set-free} music evaluation method built upon a Flow Matching backbone. In a nutshell, \method performs Flow Matching inversion on groups of latents and measures the divergence from the prior distribution. This design scheme allows divergence evaluation in a natively gaussian space, as opposed to operating in the non-gaussian latent space, as done by prior work such as FAD. We perform an extensive empirical evaluation of our methodology, reported in Section \ref{sec:results}, using objective metrics as well as through a dedicated human study. We show that \method is able to detect synthetic distortions such as white noise or low-pass filtering with high sensitivity. Our human evaluation results show that \method is highly correlated with human perception of local temporal flaws in the musical signal, demonstrated with the proposed \textit{crop-and-paste} distortion defined in section \ref{sec:results}, for which \method achieves a Pearson correlation r of $0.73$, as opposed to FAD, for which the same coefficients are $-0.87$ or $0.43$ depending on the background set. 

Furthermore, we show that \method is able to rank music generative models by their perceptual quality, without using any background set. We believe this setup is highly practical for obtaining robust automatic pipelines for assessing music generation quality.

\section{Related Work}
\label{sec:prior}
FAD \cite{kilgourfrechetaudiodistancemetric:19} is a reference-free audio quality metric commonly used to evaluate the quality of music produced by generative models. It has been widely adopted in recent years by works such as MusicLM \cite{agostinellimusiclmgeneratingmusictext:23}, MusicGen \cite{copetsimplecontrollablemusicgeneration:24}, and many others. Given a set of samples to evaluate, FAD uses a VGGish \cite{hersheycnnarchitectureslargescaleaudio:17} backbone to extract an audio feature vector for each sample, fits an empirical multivariate Gaussian distribution to the resulting embeddings, and compares it to an empirical Gaussian fitted to a large collection of studio-quality recordings, referred to as the \textit{background set}.

Recent work has explored different backbone and background set configurations for FAD \cite{guiadaptingfrechetaudiodistance:24, huangaligningtexttomusicevaluationhuman:25}. A key finding is that joint audio-text embedding models such as CLAP \cite{wulargescalecontrastivelanguageaudiopretraining:24} outperform discriminative backbones such as VGGish. The authors of MAD \cite{huangaligningtexttomusicevaluationhuman:25} also proposed replacing the Fréchet Distance with MAUVE \cite{pillutlamauvemeasuringgapneural:21} as the divergence measure.

Another recent work, KAD \cite{chungkadfadeffectiveefficient:25}, identifies two limitations of FAD: (i) its slow convergence as a function of the evaluation set size, and (ii) the strong Gaussian assumption imposed on the embedding distribution. To address these issues, KAD replaces the Gaussian fitting step of FAD with a Maximum Mean Discrepancy (MMD) objective using RBF kernels.

Our proposed methodology does not address the first limitation of FAD. However, it naturally addresses the second by measuring distances in the prior space of a Flow Matching model, whose distribution is Gaussian by construction. Finally, we note that all of the approaches discussed above - FAD, MAD, and KAD - rely on the availability of a large background reference set of clean audio, whereas our method completely eliminates the need for a background set.

\section{Method}
\label{sec:method}
We propose \method, a metric that leverages a pre-trained Flow Matching backbone to evaluate the perceptual quality of music. Given a set of encoded musical samples, \method asks the following question: \textit{How close is the inverted sample distribution to the prior?}

Specifically, we first encode a set of musical samples into a latent space and perform unconditional Flow Matching inversion to map the latent representations back to the model's prior space. We then compute the Fréchet Distance (FD) between the distribution of the transformed latents and the true prior distribution. Intuitively, if the input samples are well aligned with the data distribution on which the Flow Matching backbone was trained, the inverted latents should closely follow the prior distribution, resulting in a low FD. Conversely, as the input distribution deviates from the training distribution, the FD increases. We exploit this relationship to define our quality metric. Since the Flow Matching backbone is trained on a large corpus of high-quality, human-produced instrumental music \cite{taljointaudiosymbolicconditioning:24, copetsimplecontrollablemusicgeneration:24}, we interpret \method as a measure of music quality.

Following prior work \cite{kilgourfrechetaudiodistancemetric:19}, we model the transformed latent distribution as an empirical Gaussian. For Flow Matching inversion, we perform $100$ integration steps using Euler's method. The complete procedure for \method is described in Algorithm~\ref{alg:inv_flow_fd}.

\begin{algorithm}[ht!]
\caption{\method}
\label{alg:inv_flow_fd}
\begin{algorithmic}[1] 
    \Require a set of audio samples $\mathcal{A}$, a trained flow-matching music generation model 
    $\mathcal{M}$. \\

    $\mathcal{Z}_0 \gets \emptyset$
    \State \emph{Step 1: Flow Matching inversion:}
    \For{$a\in \mathcal{A}$}
        \State $z_{1} \gets \text{encode(a)}$
        \State $z \gets z_{1}$
        \For{$t\in 1, 0.99, 0.98, ..., 0.01$}
            \State Unconditional backward Euler step:
            \State $z \gets z - 0.01 \cdot \mathcal{M}(z, t | \emptyset)$ 
        \EndFor
    \State $z_0 \gets z$ 
    \State $\mathcal{Z}_0 \gets \mathcal{Z}_0 \cup \{z_0\}$
    \EndFor
    \\
    \State \emph{Step 2: Empirical mean and covariance estimation:}
    \State \emph{[taken over both batch and temporal dims]}
    \State $\mu \gets \text{mean}(\mathcal{Z}_0) \in \mathbb{R}^{128}$
    \State $\Sigma \gets \text{empirical-cov}(\mathcal{Z}_0) \in \mathbb{R}^{128\times128}$
    \\
    \State \emph{Step 3: Frechet Distance w.r.t. the $\mathcal{N}(0,I)$ prior:}
    \State $\textbf{FD}(\mathcal{A}|\mathcal{M}) \gets \| \mu \|^2 +tr(I + \Sigma - 2 \sqrt{\Sigma})$
    \State \Return $\textbf{FD}(\mathcal{A}|\mathcal{M})$
\end{algorithmic}
\end{algorithm}
\section{Experimental Setup}
\label{sec:exp_setup}

\textbf{Flow-Matching Backbone.} We use \textit{JASCO-400M-chords-drums}\footnote{https://huggingface.co/facebook/jasco-chords-drums-400M} \cite{taljointaudiosymbolicconditioning:24} as our Flow Matching music generation backbone, denoted by $\mathcal{M}$. JASCO is a latent Flow Matching model that operates on the continuous latent representation produced by EnCodec\footnote{https://huggingface.co/facebook/encodec\_32khz} \cite{defossezhighfidelityneuralaudio:22}, whose latent representation has a frame rate of $50$ Hz and $128$ channels. \textit{JASCO-400M-chords-drums} was originally trained to generate $10$-second music samples conditioned on text prompts, chord progressions, and drum stems. Since the model was trained with independent dropout for each conditioning modality, it can also be used as an unconditional Flow Matching model, mapping samples from the $\mathcal{N}(0, I)$ prior distribution to EnCodec latent matrices $z_1 \in \mathbb{R}^{500 \times 128}$. In this work, we use JASCO solely as an unconditional Flow Matching model trained on $\sim20k$ hours of music, as described in \cite{taljointaudiosymbolicconditioning:24}.

\textbf{Evaluation Dataset.} We use the MTG-Jamendo \cite{bogdanovmtg:19} and FMA-small \cite{defferrardfma:17} datasets for all empirical evaluations.

\textbf{FAD Measurements.} We use CLAP to extract audio embeddings for FAD computation, following prior work showing that CLAP-based FAD is highly correlated with human judgments of audio quality \cite{wulargescalecontrastivelanguageaudiopretraining:24}.

\section{Results}
\label{sec:results}

\subsection{Synthetic Distortions}\label{results:synthetic_distortions}
For all artificial distortion experiments, we use a set of $100$ randomly sampled songs from the test set of the MTG Jamendo \cite{bogdanovmtg:19}. For each configuration, we randomly crop $10$ seconds of each song, and use the set of $100$ crops as $\mathcal{A}$ for \method evaluation. In the following section we describe a set of experiments with artificial distortions, three common DSP filters, as well as a novel distortion proposed for reproducing local flaws commonly observed in generative models outputs. We use the same synthetic distortions setup for both objective and subjective evaluation.  

\textbf{White Noise.} We experiment with adding white noise $N(0, \sigma^2I)$ to the $10$ second crops, with standard deviations $5e-5, 1e-4, 5e-4, 1e-3, 2.5e-3, 5e-3, 7.5e-3, 1e-2$. Results, presented in Figure \ref{fig:inv_flow_all}a, demonstrate that our proposed metric monotonically increases with $\sigma$, and \textbf{\textit{can distinguish between levels of white noise with a sensitivity of $5e-5$ to changes in standard deviation}}. The trend is similar for FAD with the FMApop  \cite{defferrardfma:17} background set, but on the other hand, FAD with the MusicCaps \cite{agostinellimusiclmgeneratingmusictext:23} background set is unable to correctly detect mild levels of white noise.

\begin{figure*}[t!]
    \centering
    \includegraphics[scale=0.45]{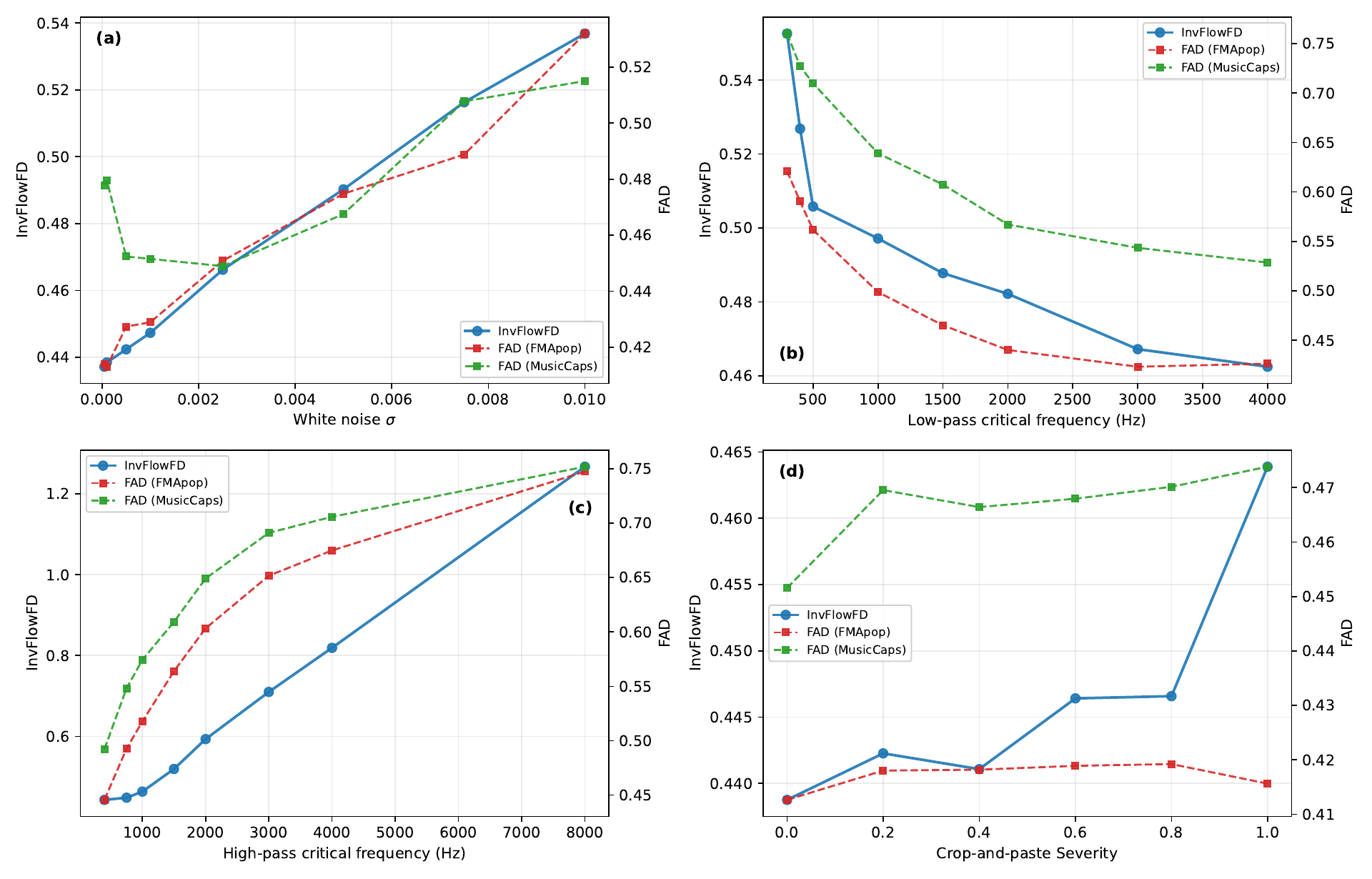} 
    \caption{\method vs. FAD as a function of different distortion parameters.\label{fig:inv_flow_all}}
\end{figure*}

\textbf{Low Pass Filter.} We experiment with applying low pass filter to the samples, with critical frequencies [Hz] of $300, 400, 500, 1k, 1.5k, 2k, 3k \text{ and } 4k$. We use the torchaudio implementation from \texttt{lowpass\_biquad}. Results, reported in Figure \ref{fig:inv_flow_all}b, demonstrate a monotonical decrease of \method as function of the low pass critical frequency. FAD with both background sets behaves similarly.

\textbf{High Pass Filter.} Similarly, we apply high pass filter to the samples, with critical frequencies [Hz]  of $400, 750, 1k, 1.5k, 2k, 3k, 4k, \text{ and } 8k$. We use the implementation from torchaudio: \texttt{highpass\_biquad}. Results, reported in Figure \ref{fig:inv_flow_all}c, demonstrate a monotonical increase of \method as function of the high pass critical frequency. FAD, with both background sets, reacts similarly. 

\textbf{Crop-and-paste.} We simulate mispredictions of audio frames in the time domain by copying and overlaying filtered fractions of the original sample. The samples are either cut from $300$Hz to $4$kHz, or from $4$kHz to $10$kHz, with the same probability for every frequency to be chosen for every cut. A single ``severity'' argument dictates three parameters simultaneously: (i) The amplitude multiplier of the injected patch (scaled up to a factor of $1.6$, which intentionally triggers a hard digital clipper); (ii) The density of the artifacts (triggering glitch events on up to $4\%$ of the STFT time frames) and; (iii) The temporal duration of each copied patch (randomly sampled from boundaries linearly proportional to the severity). This operation targets mid or high-frequency bands, to try and mimic the miss-prediction of human-played instruments. By doing this, we try to break the coherent time-structure of music, and not just distort it locally. If a metric will manage to detect such time-related alterations of human played audio, we can say that it evaluates not only how music should sound, but also how elements playing should be structured. 

We apply the crop-and-paste distortion with severity levels in $
\{0.0, 0.2, 0.4, 0.6, 0.8, 1.0\}$. Results, reported in Figure \ref{fig:inv_flow_all}d, show an increase of \method as function of the the crop-and-paste severity, suggesting \method is able to successfully detect local flaws in music coherency. For FAD, on the other hand, the trend is non monotonic, specifically for the FMA-pop background set.

\begin{figure}[t!]
    \centering
    \includegraphics[scale=0.45]{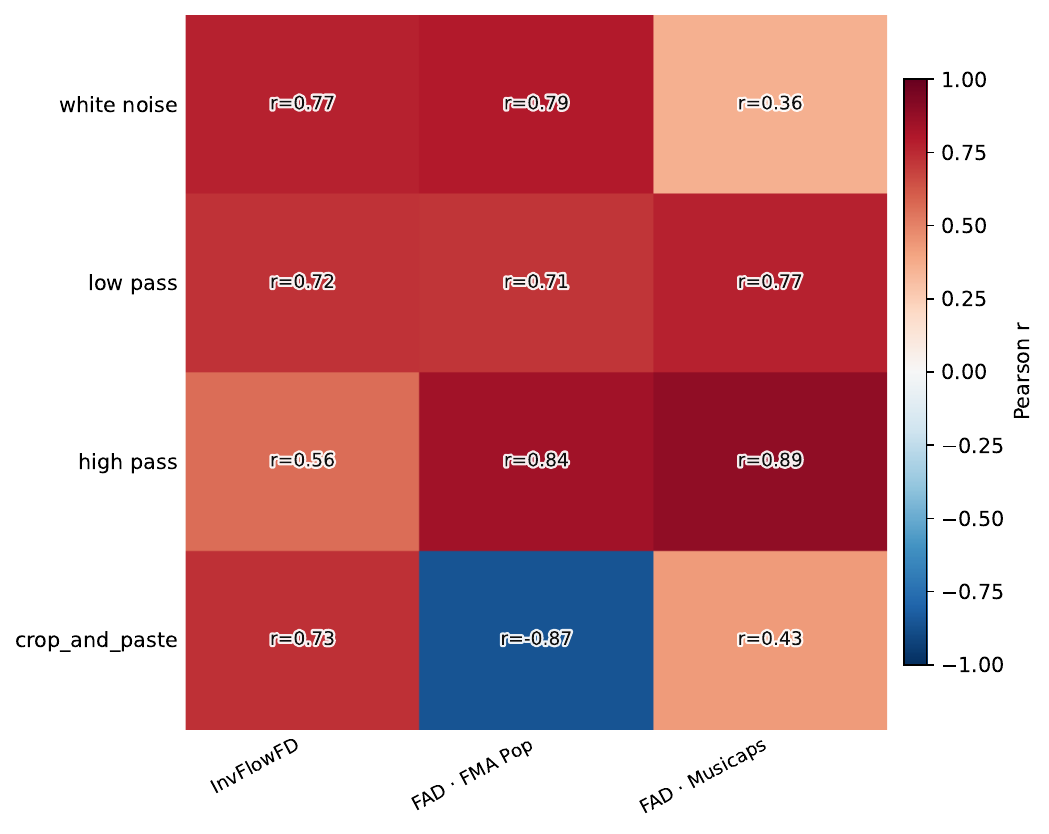}
    \caption{Pearson correlation between automatic metrics and human-based worth.}
    \label{fig:humaneval_correlations}
\end{figure}

\subsection{Human Study}
We conduct a pairwise human preference evaluation using a locally hosted Gradio \cite{abidgradiohasslefreesharingtesting:19} web interface. On each trial, the interface presents two $10$s MP3 excerpts (A/B) and ask the rater, similar to the human evaluation setup in \cite{kilgourfrechetaudiodistancemetric:19}, to \textit{select the sample that sounds cleaner and more professionally produced}. Stimuli are drawn from an augmented version of $100$ random samples from Jamendo \cite{bogdanovmtg:19} test set. The augmentations we use are the same as in previous section, namely white noise, low-pass filtering, high-pass filtering and our proposed 'crop-and-paste' distortion. We use the exact same critical parameter values described in the previous section. Additionally, we use the clean samples as pseudo-distortions. For each trial we sample an augmentation type uniformly at random from \textit{$\{\text{white noise}, \text{low-pass}, \text{high-pass}, \text{crop-and-paste}\}$}, then select two different distortion levels at random. We use aligned A/B samples that differ only in augmentation level; A/B placement is randomized to mitigate position bias. 

\begin{figure*}[t!]
    \centering
    \hspace*{0.1cm}
    \includegraphics[scale=0.64]{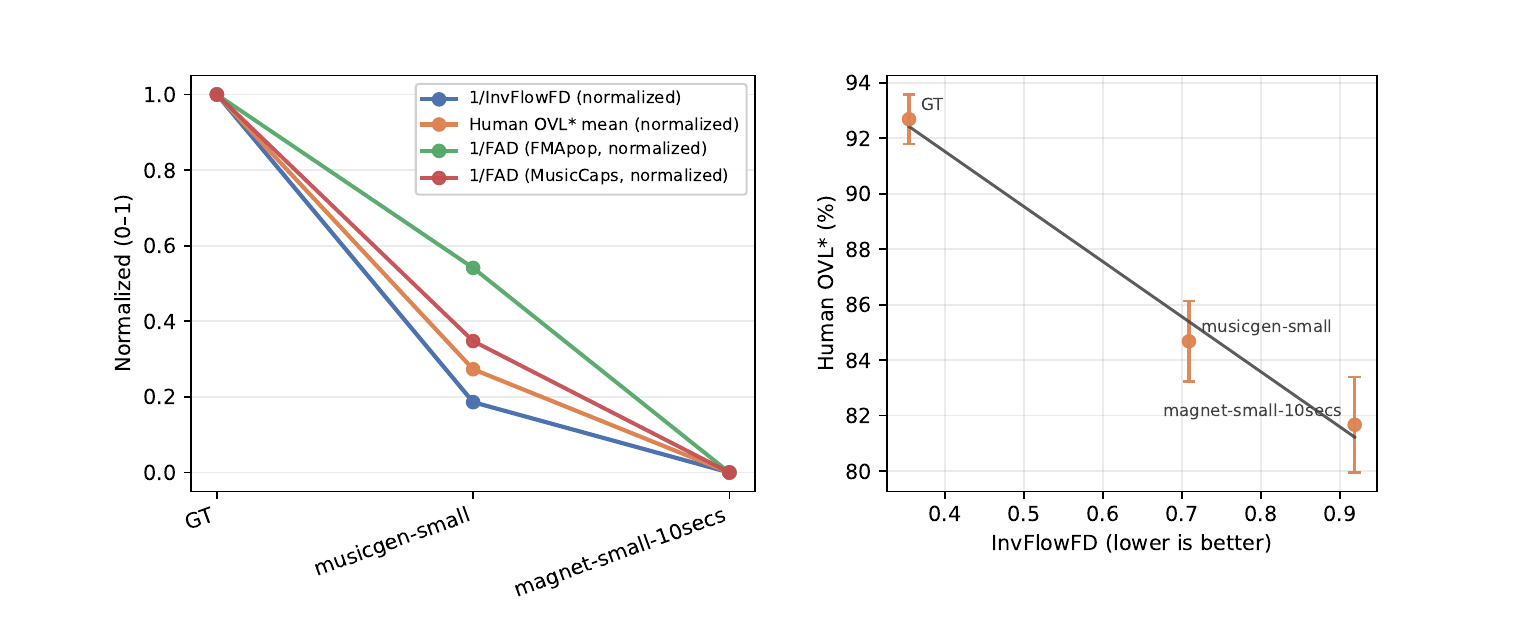} 
    \caption{Left: \method and FAD evaluated on generative music models and compared to human evaluation results from \cite{zivmaskedaudiogenerationusing:24}$^*$. Right: Correlation plot of human evaluation overall quality measured from MAGNeT \cite{zivmaskedaudiogenerationusing:24}$^*$ and \method, demonstrating strong correlation.}
    \label{fig:fd_per_generative_model}
\end{figure*}

\begin{figure}[t!]    
    \centering
    \includegraphics[scale=0.25]{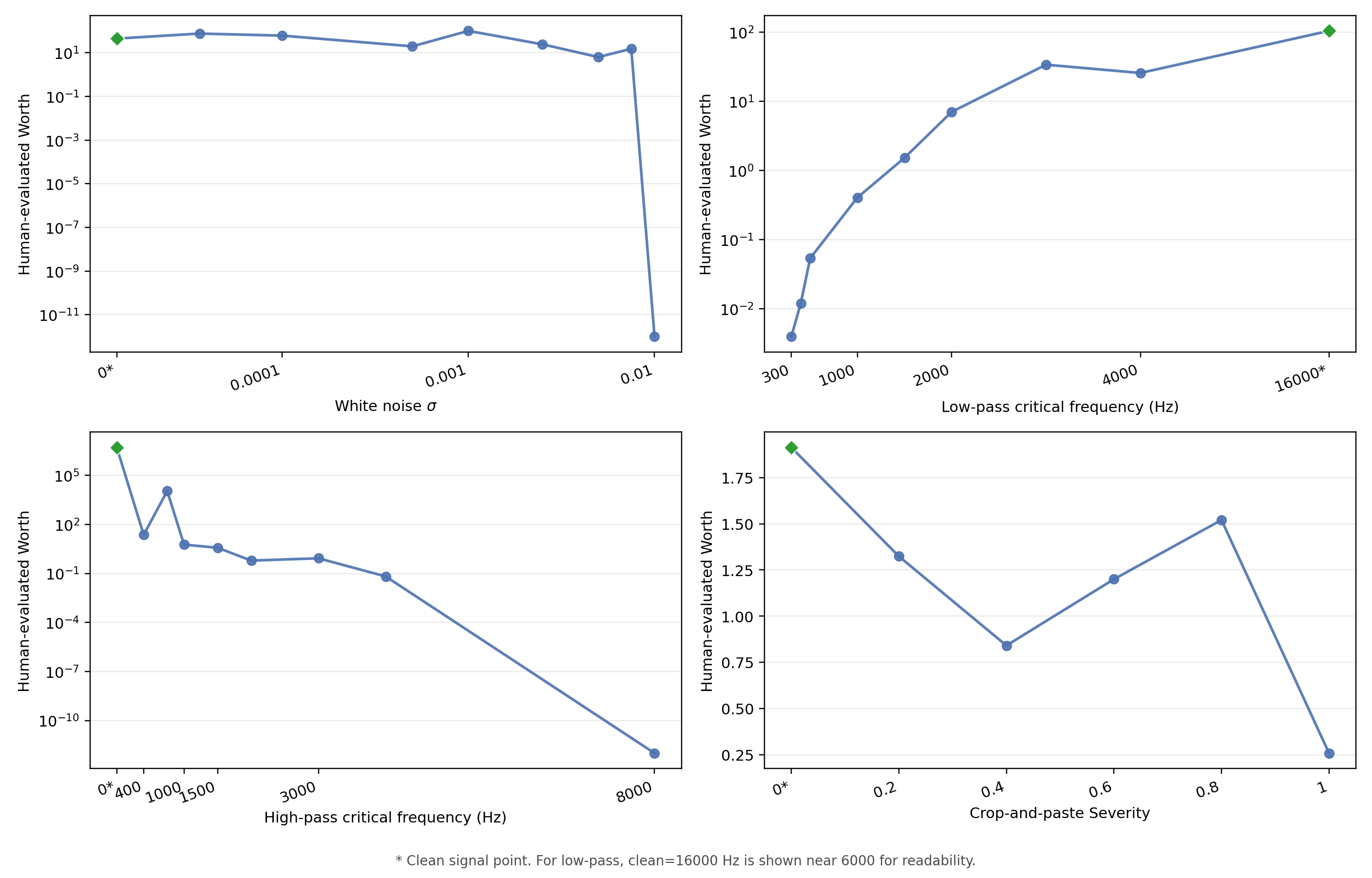}
    \caption{Human Evaluation Result: Worth curves for synthetic distortions.}
    \label{fig:humaneval_worth_curves}
\end{figure}

We collect a total of $400$ pairwise responses, from $8$ different human raters, resulting in an average of $3$-$5$ responses per pair of distortion levels. We then estimate worth values by applying the Plackett-Luce \cite{luceindividual:1959} algorithm per each type of distortion. We presented the worth values per distortion family in Figure \ref{fig:humaneval_worth_curves}. Finally, we measure the Pearson correlation between each automatic metric, including the baseline FAD, and report the results in Figure \ref{fig:humaneval_correlations}. For clarity, we flip the metrics sign to obtain positive correlations.

Results demonstrate the ambiguity of FAD, for which correlation changes as function of the background set (here we experimented with FAD with two background sets - FMA-pop and MusicCaps), specifically for white noise. Results also show that \method is on par with FAD in terms of correlation to human perception of low-pass, and white noise, but slightly worse than FAD in high-pass correlation. For local, subtle distortions such as the proposed crop-and-paste function, \method is much more correlated with human perception, achieving a Pearson correlation coefficient r of $0.73$, while the same coefficients for FAD are $-0.87$ and $0.43$ for the FMA-pop and MusicCaps background sets respectively. 

\subsection{Comparing Music Generation Models with \method}
We demonstrate how \method could be used for automatic, reference-free and \textbf{\textit{background-set-free}} evaluation of the quality of samples drawn from generative music models.
We evaluate each model by sampling $100$ text-to-music samples with prompts defined by the tags of the test set of MTG Jamendo dataset. The tags are deterministically converted to textual descriptions such as \textit{"alternative track with electronic experimental vibes."} or \textit{"indie track with newwave vibes with flute, with dream epic mood."}. 
In Figure \ref{fig:fd_per_generative_model} we report \method and FAD, for two text-to-music generation models taken from prior work \cite{copetsimplecontrollablemusicgeneration:24, zivmaskedaudiogenerationusing:24}. 

In addition, we plot the human evaluation results from MAGNeT \cite{zivmaskedaudiogenerationusing:24} for overall quality, and we mark it as OVL$^{*}$ in both graphs of Figure \ref{fig:fd_per_generative_model}, in order to demonstrate the correlation of our proposed method with human opinion. The reference human evaluation results suggest that the perceptual quality of 
\textit{musicgen-small}~\cite{copetsimplecontrollablemusicgeneration:24} is better than the perceptual quality of \textit{magnet-small-$10$secs}~\cite{zivmaskedaudiogenerationusing:24}, and that the perceptual quality of the ground truth MTG Jamendo songs is better than the quality of the samples generated by both generative models. 

Results for both FAD and \method demonstrate a similar trend, reassuring the ability of FAD to rank generative models, and more importantly, showing that the proposed \method is capable of correctly ranking generative models without requiring neither a reference set nor a background set.

\begin{figure*}[t!]    
    \centering
    \includegraphics[scale=0.42]{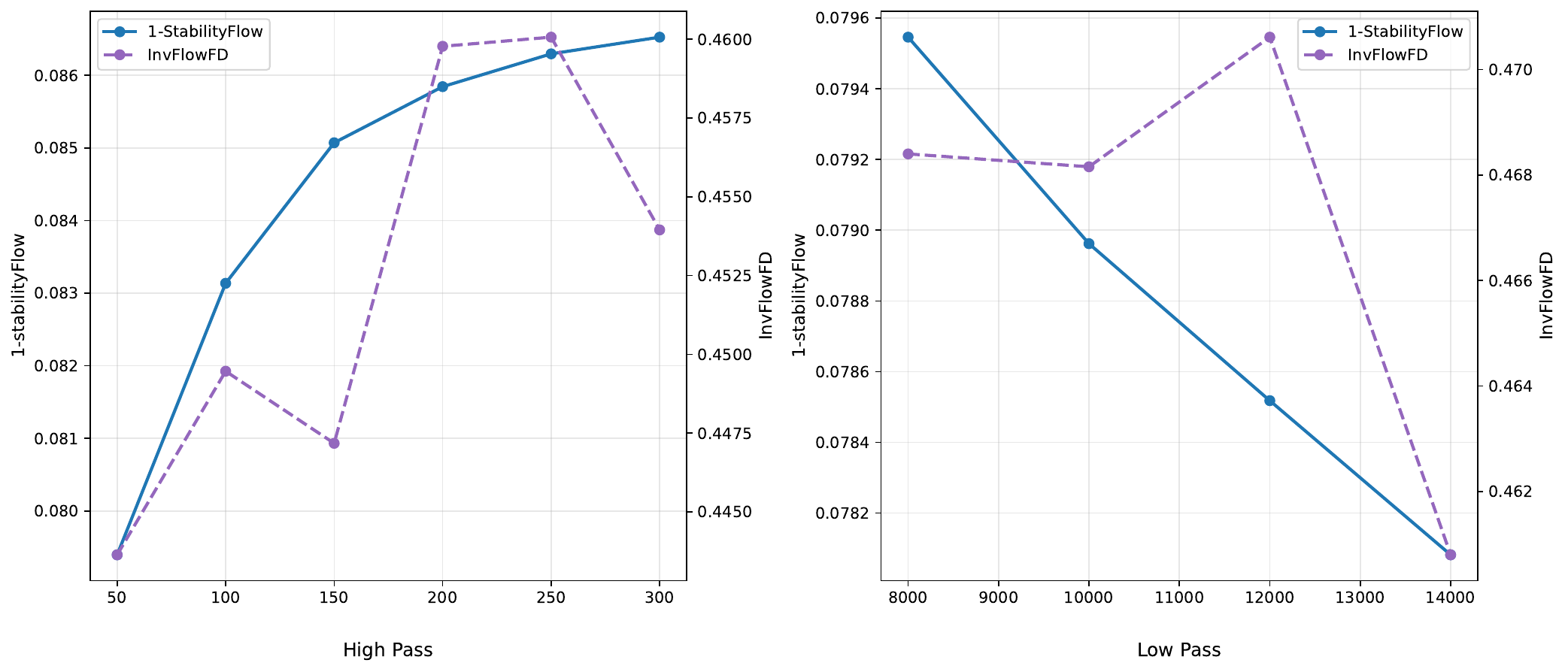}
        \caption{Evaluation of \stabilityMethod and \method on subtle synthetic distortions. Note that we report $1-\stabilityMethod$ for clarity, ensuring that both metrics go up as function of the distortion severeness. }
    \label{fig:stability_vs_invflow}
\end{figure*}

\section{Analysis}
\label{sec:analysis}

Another demonstration of the robustness of Flow Matching for music-quality evaluation concerns the following question: \textit{Given a single sample, can a pretrained flow-matching model determine how closely that sample resembles its training data?} To address this question, we propose \stabilityMethod.

\stabilityMethod is a way to use a Flow Matching generative model as an evaluator, by measuring the stability of a local back-and-forth flow transformation. For in-distribution samples of audio, we expect the latents encoded by EnCodec to remain stable under such a transformation. For out-of-distribution samples, we expect the reconstructed latents to be different from the original ones - as the model's vector field will push the out-of-distribution latents to some other in-distribution point. 

Practically, we perform the following process: For a step size $s$ and a flow-matching model $\mathcal{M}$, we first invert an encoded latent sample $z$ by taking a single large Euler step toward the prior, computing $z_{noisy} = z - s \cdot \mathcal{M}(z)$. Then, we perform $k$ forward Euler steps to reconstruct $\hat{z}$. 
Finally, we compute the cosine similarity between each sample and its reconstructed version - which is another key factor of the method, as it is a per-sample metric. In our experiments, we set $s=0.7, k=10$ and $\mathcal{M}$ is \textit{JASCO-400M-chords-drums}. The full process is described in Algorithm \ref{alg:reconstruction_ed}.

\begin{algorithm}[t!]
\caption{\stabilityMethod}
\label{alg:reconstruction_ed}
\begin{algorithmic}[1] 
    
    \Require an ordered sequence of audio samples $\mathcal{A} = (a_1, \dots, a_N)$, Trained flow-matching model $\mathcal{M}$, step size $s$, number of steps to reconstruct $n$.

    \State $\mathcal{Z} \gets ()$ \Comment{Sequence of original latents}
    \State $\hat{\mathcal{Z}} \gets ()$ \Comment{Sequence of reconstructed latents}
    \State $t \gets 0.999$ \Comment{Initial time parameter for inversion}
    
    \For{$i \gets 1$ \textbf{to} $N$}
        \State $z_i \gets \text{encode}(a_i)$
        
        \State \emph{Step 1: Single large inversion step}
        \State $z_{noisy} \gets z_i - s \cdot \mathcal{M}(z_i, t | \emptyset)$
        \State $t \gets t - s$
        
        \State \emph{Step 2: n-step reconstruction}
        \State $\hat{z}_i \gets z_{noisy}$
        \State $\Delta t \gets s / n$
        \For{$j \gets 1$ \textbf{to} $n$}
            \State \emph{Forward Euler step:}
            \State $\hat{z}_i \gets \hat{z}_i + \Delta t \cdot \mathcal{M}(\hat{z}_i, t | \emptyset)$
            \State $t \gets t + \Delta t$
        \EndFor
        
        \State $\mathcal{Z}[i] \gets z_i$
        \State $\hat{\mathcal{Z}}[i] \gets \hat{z}_i$
    \EndFor

\State \emph{Step 3: Average Pairwise Cosine Similarity}
    \State $\textbf{CS}(\mathcal{A}|\mathcal{M}, s) \gets \frac{1}{N} \sum_{i=1}^{N} \frac{z_i \cdot \hat{z}_i}{\|z_i\|_2 \|\hat{z}_i\|_2}$
    \State \Return $\textbf{CS}(\mathcal{A}|\mathcal{M}, s)$
\end{algorithmic}
\end{algorithm}


Although both \stabilityMethod and \method evaluate music quality, they offer complementary advantages. \method directly compares the distribution of a set of inverted samples with the prior distribution, making it well suited for assessing distribution-level similarity to the backbone model's training data. In contrast, \stabilityMethod performs multiple forward passes through the backbone and uses pairwise cosine similarity. Therefore, it supports sample-level quality assessment.

We also found that \stabilityMethod was more sensitive to some subtle distortions that were not captured by \method. To examine this sensitivity, we applied a high-pass filter with cutoff frequencies ranging from $50$ Hz to $300$ Hz and a low-pass filter ranging from $8000$ Hz to $14000$ Hz. The results, presented in Figure~\ref{fig:stability_vs_invflow}, suggest that \stabilityMethod is more sensitive to subtle distortions. For example, low pass 10kHz was ranked as better than low pass 12kHz by \method, while \stabilityMethod presents a monotonic reaction. Performance on these subtler distortions suggests that \stabilityMethod may be particularly useful in settings that prioritize precise sample-level assessment over distribution-level evaluation.
These measurements were obtained using the same samples used to evaluate \method in figure \ref{fig:inv_flow_all}, but with the subtler cutoff frequencies described above.

\section{Conclusion}
\label{sec:conclusion}

We introduced \method, a background-set-free metric for evaluating the perceptual quality of music using a pre-trained Flow Matching backbone. Unlike prior approaches such as FAD, \method does not rely on curated reference datasets, eliminating a major source of evaluation bias. Through synthetic distortions, human studies, and evaluation of generative music models, we demonstrated that \method is sensitive to perceptual degradations, correlates well with human judgments, and provides a robust alternative to background-dependent metrics. We further showed that Flow Matching models can serve as intrinsic evaluators of music quality, and introduced \stabilityMethod as a complementary sample-level metric. 

Beyond evaluation, we believe this perspective opens new opportunities for using generative backbones as reward models for optimizing larger music generation systems. More broadly, we hope this work encourages viewing generative models not only as synthesizers, but also as perceptual evaluators that can enable more practical and robust evaluation pipelines for music generation.

\paragraph{Acknowledgments.} This research work was supported by Israel Science Foundation (ISF), grant number $2049/22$. 







\bibliography{refs}

%
%
%
%

\end{document}